\documentclass[12pt]{article}
\usepackage{epsfig,amssymb}
\usepackage{latexsym}

\newcommand{\bq}{\begin{equation}}
\newcommand{\eq}{\end{equation}}
\newcommand{\bqa}{\begin{eqnarray}}
\newcommand{\eqa}{\end{eqnarray}}
\newcommand{\ben}{\begin{enumerate}}
\newcommand{\een}{\end{enumerate}}
\newcommand{\bc}{\begin{center}}
\newcommand{\ec}{\end{center}}
\newcommand{\bqb}{\begin{eqnarray*}}
\newcommand{\eqb}{\end{eqnarray*}}

\def\gsim{\gtrsim}

\def\pr#1#2#3{ Phys. Rev. ${\bf{#1}}$:#2 (#3)}
\def\prl#1#2#3{ Phys. Rev. Lett. ${\bf{#1}}$:#2 (#3)}

\def\epj#1#2#3{ Eur. Phys. J. ${\bf{#1}}$:#2 (#3)}

\global\nulldelimiterspace = 0pt

\def\mwd{m_W^2}

\def\mzd{m_Z^2}

\def\tchi{\tilde \chi}

\def\mchi{m_{\tchi}}
\def\mdt{m_{\tilde d_L}}

\begin{document}

\begin{flushright}
January  2010\\
%arXiv: xxxxxxx  [hep-ph]\\
\end{flushright}

\vspace{2cm}
%---------------------titre ---------------------------------------
\begin{center}
{\Large {\bf Asymptotic Helicity Conservation in SUSY  } } \\
 \vspace{0.5cm}
%-----------------------------------------------------------------
{\bf G.J. Gounaris$^a$, J. Layssac$^b$,
and F.M. Renard$^b$}\\
\vspace{0.2cm}
$^a$Department of Theoretical Physics, Aristotle
University of Thessaloniki,\\
Gr-54124, Thessaloniki, Greece.\\

$^b$Laboratoire de Physique Th\'{e}orique et Astroparticules,
UMR 5207\\
Universit\'{e} Montpellier II,
 F-34095 Montpellier Cedex 5.\\

\vspace{0.2cm}
 Talk presented by GJG at the meeting  on the "Standard Model and beyond and
 the Standard Cosmology", held at  Corfu (Greece), during August 30 - September 6, 2009.
\end{center}

\vspace*{0.5cm}
\begin{center}
{\bf Abstract}
\end{center}

We summarize the extensive work started in \cite{heli}, according to which
total helicity is conserved for any two-to-two process,
at   $\sqrt{s} \gg M_{SUSY}$ and fixed angles,  in any   SUSY extension of SM.
Asymptotically the theorem is exact. But   it may also have important implications
at lower energies $\sqrt{s} \gsim  M_{SUSY}$. Up to now, these have been investigated
to 1loop electroweak (EW) order for the processes  $ug \to d W^+ ,  \tilde d_L \tchi^+_i$;
as well as     the 17  $gg\to HH'$, and the 9  $gg\to VH$ processes,
where $H,H'$ denote  Higgs   or Goldstone bosons, and  $V= Z, W^\pm$.

\def\thefootnote{\arabic{footnote}}
\setcounter{footnote}{0}

\vspace{2cm}

Some years ago it has been established, to all orders in MSSM  perturbation,
 that for any two-to-two process
 \bq
a_{\lambda_a}+b_{\lambda_b} \to c_{\lambda_c}+d_{\lambda_d} ~~, \label{gen-process}
\eq
where   $\lambda_j$ denotes  the particle helicity, all amplitudes satisfying
\bq
\lambda_a+\lambda_b -\lambda_c-\lambda_d \neq 0 ~~, \label{HV-constraint}
\eq
vanish exactly at asymptotic energies and fixed angle  \cite{heli}.
This property, which is  also true  to any non-minimal supersymmetric
extension of SM,  has been termed    asymptotic  Helicity Conservation (HCns) \cite{gghhVh}.
The amplitudes obeying  (\ref{HV-constraint}),
are called below  helicity violating (HV) amplitudes; while those
satisfying  $\lambda_a+\lambda_b -\lambda_c-\lambda_d= 0$,
which are the only ones that can be non-vanishing asymptotically,
are termed as helicity conserving (HC) amplitudes.

HCns is an impressive SUSY property, drastically reducing  the number
of the asymptotically non-vanishing amplitudes, independently of the softly breaking sector.
Because of this,  it may well be considered
at the same level  as the best    SUSY beauties,  like the  smooth  ultraviolet  behavior,
 the   gauge coupling unification and  existence   of  dark matter candidates \cite{gghhVh}.

If no external vector bosons appear in (\ref{gen-process}), HCns is merely a consequence
of the chiral structure of the fields in SUSY.
As a result, it  is valid at a diagram-by-diagram level,
to all orders in perturbation theory.

When external gauge bosons are involved in (\ref{gen-process}) though,
\underline{huge} cancelations among the various  diagrams are essential for
establishing  HCns \cite{heli}.  This may easily be seen at the Born level, but it should much
more impressive at higher order \cite{heli}. In constructing the  general proof in this case,
a crucial  role was played by the fact that
the SUSY  transformation for  $"gauge \leftrightarrow gaugino"$, when projected
to single particle states, only involves particles with helicities of the same sign \cite{heli}.
Thus, the structure of the amplitudes involving  external gauginos, was used
to study those involving external gauge bosons.\\

As a first  example
we quote the complete   1loop calculations of the   processes
$\gamma\gamma \to \gamma \gamma$, $\gamma\gamma \to \gamma Z$ or
$\gamma\gamma \to ZZ$, where all HV amplitudes obeying (\ref{HV-constraint})
were found  to tend asymptotically to non-vanishing constants in SM,
while vanishing in MSSM \cite{gamgamV10V20}. Thus, the SM  and the sfermion-loop contributions
to all HV amplitudes for these processes go asymptotically to opposite constants,
exactly canceling  each other in SUSY  \cite{gamgamV10V20}.

Moreover, in  the  1loop  study of
 $\gamma\gamma \to \gamma \gamma,~ \gamma Z,~  ZZ$   \cite{gamgamV10V20},
it has been observed that the HC amplitudes  were asymptotically much
larger that the HV ones, also in SM.  Thus,  HCns  for these processes,
appears approximately true in SM also.
A similar property is also observed for  $ug\to dW$ in \cite{ugdW}.

HCns is not a general property of  SM  though. Thus for the 1loop EW computations of
  $gg \to H^0 H^0,~ W^+ G^-, ZG^0, G^+G^- $
(where $G$  denotes goldstone bosons), we have found a strong violation of HCns in SM ;
while, of course, it  is respected in all     MSSM gluon-fusion processes
to a pair of spin=1 or spin=0 bosons \cite{gghhVh}. Such examples indicate
that HCns is a genuine SUSY property. \\

\begin{figure}[t]
\vspace*{-2.cm}
\[
\hspace{-0.5cm}
\epsfig{file=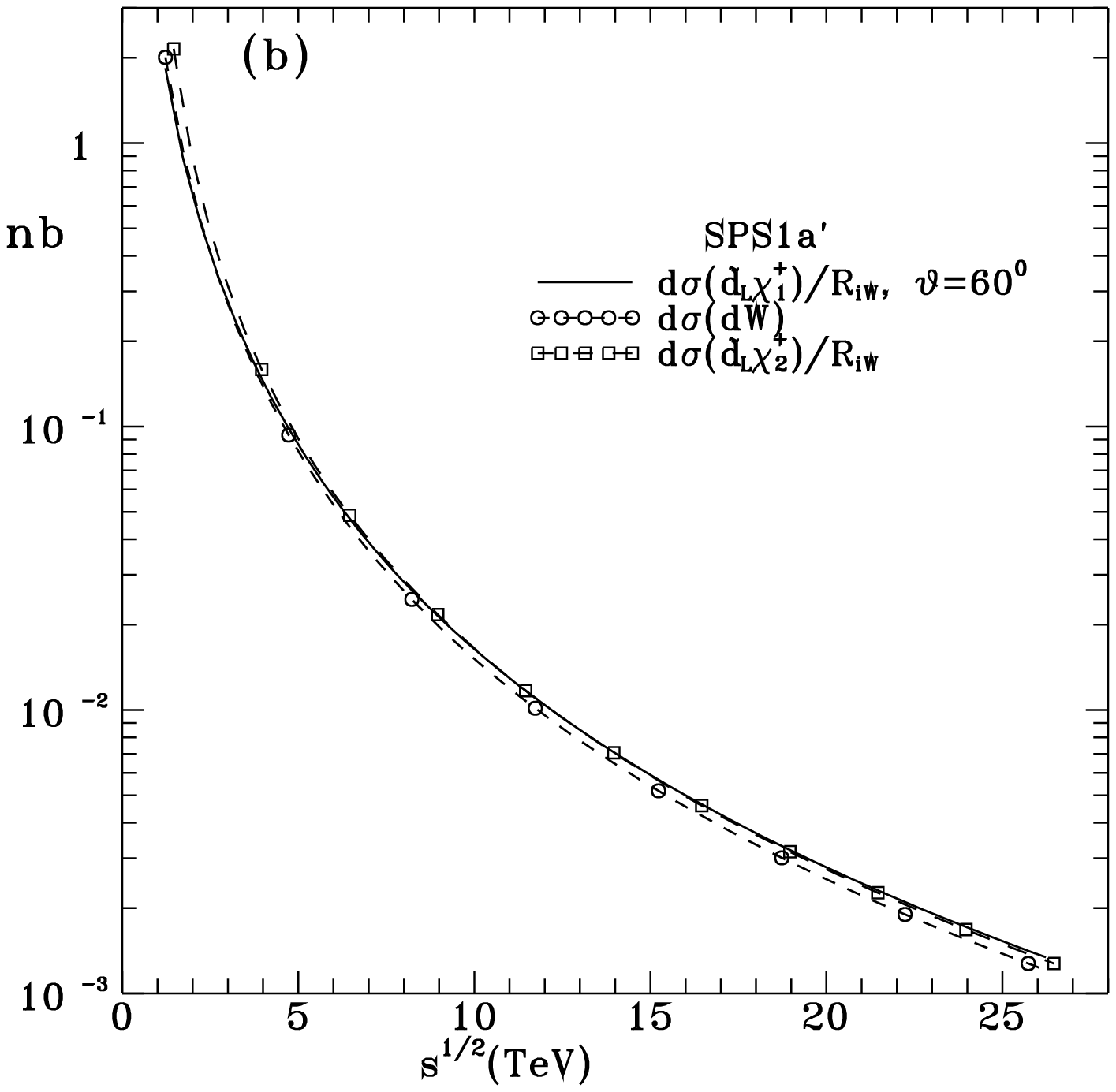,height=6.cm}\hspace{0.5cm}
\epsfig{file=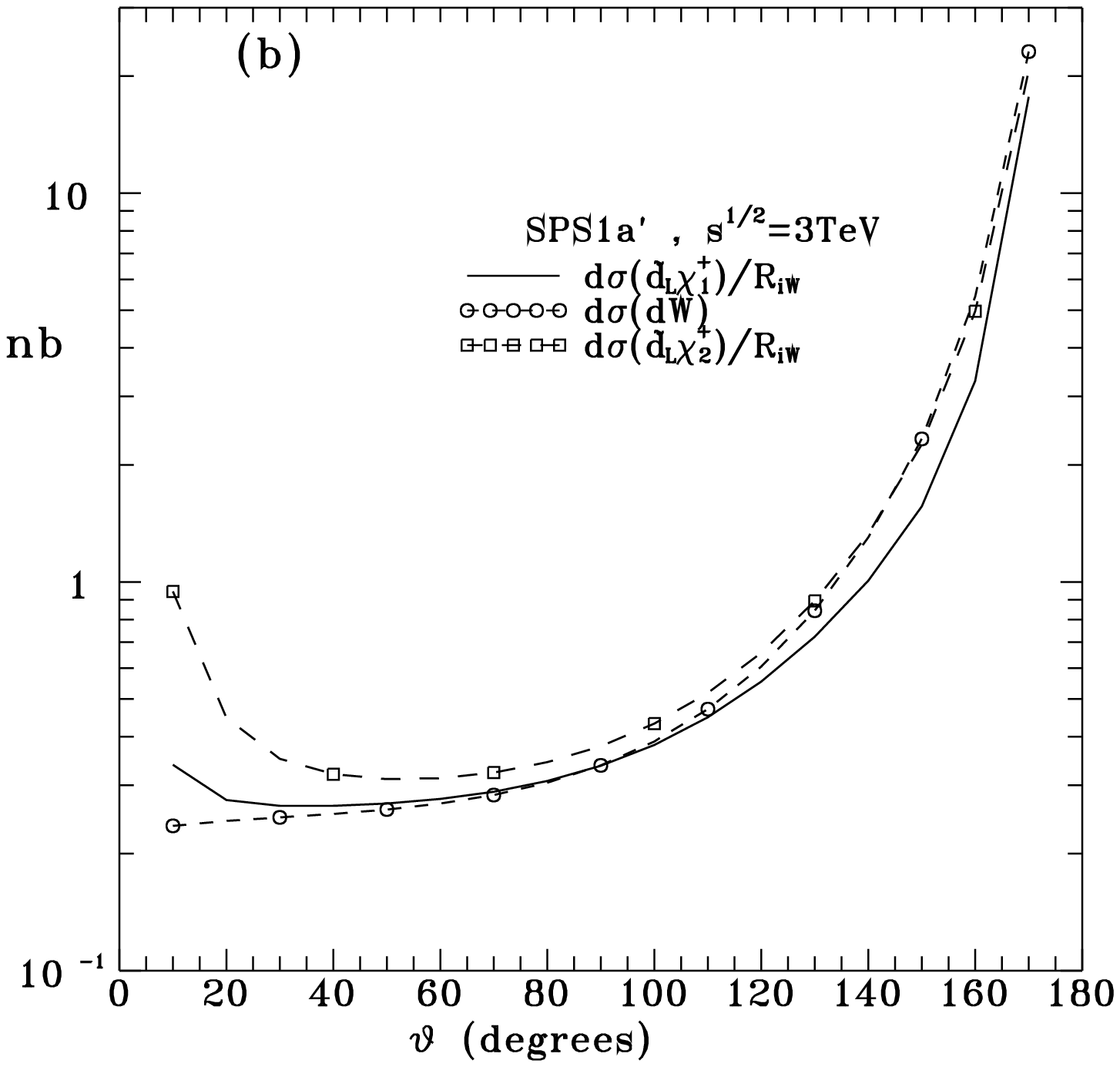,height=6.cm}
\]
\vspace*{-1.2cm}
\caption[1]{ \begin{small}
 The energy (left panel) and angular (right panel)
 dependencies of the left part
of (\ref{dW-sdchi-rel}) (dash line with circles), are compared
to the right part for     $\tchi^+_1$ (full line)
and  $\tchi^+_2$  (dash line with squares) production,
  using  the bench mark model $SPS1a'$ \cite{SPA}.
  \end{small}}
\label{rel1-fig}
\end{figure}

The  dominance of the HC amplitudes at asymptotic energies  allows the construction
of simple relations among the differential cross sections of various processes,
which become exact at high energies, but may be also useful at the  LHC range,
provided  the SUSY  scale is not too high. One such example is obtained using
\bqa
 \frac{d\hat \sigma(u g\to \tilde d_L \tchi_i^+)}
{d\cos\theta} &= &  \frac{\beta_{\tchi  }'}{3072 \pi s}
 \sum_{\lambda_u \lambda_g \lambda_{\tchi} }
 |F^{\tchi}_{\lambda_u \lambda_g \lambda_{\tchi}}|^2 ~~, \label{dsigma-chi-theta} \\
 \frac{d\hat \sigma(u g\to d W^+)}{d\cos\theta}
&= & \frac{\beta_W'}{3072 \pi s} \sum_{\lambda_u \lambda_g \lambda_d \lambda_W }
|F^{W}_{\lambda_u \lambda_g \lambda_d \lambda_W }|^2  ~~, \label{dsigma-W-theta}
\eqa
\bqa
&& \beta'_{\tchi_i}={2p'_{\tchi_i}\over\sqrt{s}}~~,~~\beta_W' = 1-\frac{\mwd}{s}~~,~~
  a_{\tchi_i W}= \frac{\alpha}{4\pi}{(1+26c^2_W)\over 72 s^2_Wc^2_W} \ln\frac{M_{SUSY}^2}{\mzd}
  ~~, \nonumber \\
&& R_{iW} = \frac{[s-(\mchi +\mdt)^2]^{1/2}[s-(\mchi -\mdt)^2]^{1/2}}{s-\mwd}
|Z^-_{1i}|^2 \frac{(1+a_{\tchi W})^2 \sin^2 \theta }{5+2\cos\theta +\cos^2\theta}
 ~~, \label{kin-dW-sdchi}
\eqa
with $p'_{\tchi_i}$ being the c.m. momentum of the produced chargino  $\tchi_i$.
HCns then implies the asymptotic relation  \cite{ug-dW-sdWino}
\bq
 \frac{d\hat \sigma(u g\to d W^+)}{d\cos\theta}
 \simeq \frac{1}{R_{iW}}
 \frac{d\hat \sigma(u g\to \tilde d_L \tchi_i^+)}{d\cos\theta} ~~.
 \label{dW-sdchi-rel}
\eq
As shown in  Fig.\ref{rel1-fig},
 this relation  is  quite accurate even at the LHC energy range,
provided  the SUSY masses are close to those of
the $SPS1a'$ benchmark \cite{SPA}. \\

In \cite{gghhVh},  the gluon-gluon fusion process to two colorless  scalars,
or a scalar and vector
were considered, which  most stringently test HCns, since they receive no Born contribution.
Many  asymptotic relations were then derived, one set of which is
\bqa
 &R_1 \Rightarrow & \tilde \sigma (gg\to G^0G^0)  \simeq \tilde \sigma (gg\to G^0A^0)
\left (\frac{R_{a1}}{R_{a2}} \right )^2
\simeq \tilde \sigma (gg\to A^0A^0) \left (\frac{R_{a1}}{R_{a3}} \right )^2 \nonumber \\
&& \simeq \tilde \sigma (gg\to H^0H^0) \left (\frac{R_{a1}}{R_{a4}} \right )^2
\simeq \tilde \sigma (gg\to h^0h^0) \left (\frac{R_{a1}}{R_{a5}} \right )^2
\simeq \tilde \sigma (gg\to H^0h^0) \left (\frac{R_{a1}}{R_{a6}} \right )^2 \nonumber \\
 && \simeq \tilde \sigma (gg\to Z^0G^0)  \simeq \tilde \sigma (gg\to Z^0A^0)
\left (\frac{R_{a1}}{R_{a2}} \right )^2 ~~,  \label{R1-rel}
\eqa
where $ \tilde \sigma(gg\to HH', VH)$ are differential cross sections from which kinematical
factors have been removed, and  $R_{aj}$ are numerical constants depending on the mixing
angles of the Higgs-sector \cite{gghhVh}.  If the SUSY masses are close
to those in $SPS1a'$, some  of the relations (\ref{R1-rel}) are approximately true even
at energies close to the LHC range. Work is in progress for extending this study
to include the processes $gg \to VV', ~\tchi_i^+\tchi_j^-,~\tchi_i^0\tchi_j^0 $.\\

In conclusion we emphasize that
 HCns is a genuine  SUSY property,  which strongly simplifies
the asymptotic 2-to-2 amplitudes. It solely depends on the symmetry; not
on its breaking! And it is  this symmetry that guarantees  the cancelation of the strong
divergencies between the fermion and the boson loops, which creates HCns  and the SUSY
beauties recapitulated at the beginning.

HCns  provides many asymptotic relations among various subprocess cross sections.
 If the SUSY scale is not too high, these  may be useful for LHC,
 or a future higher energy machine.

Codes  for the amplitudes of the 1loop EW process used in this work,
are  available in http://users.auth.gr/gounaris/ FORTRANcodes.

{\bf Acknowledgement}: GJG was partially supported by the European Union
 contract MRTN-CT-2006-035505 HEPTOOLS.


\begin{thebibliography}{99}
%
\bibitem{heli}  G.J. Gounaris and F.M. Renard,
\prl{94}{131601}{2005},  hep-ph/0501046;
Addendum in \pr{D73}{097301}{2006},  hep-ph/0604041.
%
\bibitem{gghhVh} G.J. Gounaris, J. Layssac
and F.M. Renard, \pr{D80}{013009}{2009}, arXiv:0903.4532 [hep-ph].
%
\bibitem{gamgamV10V20} G.J. Gounaris, P.I. Porfyriadis
  and F.M.Renard, \epj{C9}{673}{1999}, arXiv:hep-ph/9902230.
 G.J. Gounaris, J.Layssac, P.I. Porfyriadis   and F.M.Renard, \epj{C10}{499}{1999},
 arXiv:hep-ph/9904450.
  G.J. Gounaris, J.Layssac, P.I. Porfyriadis
  and F.M.Renard, \epj{C13}{79}{2000}, arXiv:hep-ph/9909243.
   G.J. Gounaris, P.I. Porfyriadis
  and F.M.Renard, \epj{C19}{57}{2001}, arXiv:hep-ph/00100006.
%
\bibitem{ugdW} G.J. Gounaris, J. Layssac
and F.M. Renard, \pr{D77}{013003}{2008}, arXiv:0709.1789 [hep-ph].
%
\bibitem{ug-dW-sdWino} G.J. Gounaris, J. Layssac
and F.M. Renard, \pr{D77}{093007}{2008}, arXiv:0803.0813 [hep-ph].
%
\bibitem{code}  The FORTRAN codes together with a Readme file explaining
its use, are contained in  gghhcode.tar.gz and ggVhcode.tar.gz,
which can  be downloaded from
http://users.auth.gr/gounaris/FORTRANcodes. All input parameters
in the code are at the electroweak scale.

\bibitem{SPA} J.A. Aguilar-Saavedra et al., SPA convention,
\epj{C46}{43}{2005}, hep-ph/0511344;
B.C. Allanach et al. \epj{C25}{113}{2002}, hep-ph/0202233.
%

\end{thebibliography}
\end{document}